\documentstyle[aps,multicol]{revtex}
\input {epsf.sty}
\begin{document}
\begin{multicols}{2}
\noindent {\Large \bf  Spatially resolved electronic structure inside and outside the
vortex core  of  a high temperature superconductor }
\\ \\ {\bf V. F. Mitrovi{\'c}$^\star$, E. E. Sigmund$^\star$, M.
Eschrig$^\dagger$, H. N.
Bachman$^\star$, W. P.  Halperin$^\star$, A. P. Reyes$^\ddag$, P.
Kuhns$^\ddag$ \& W. G.
Moulton$^\ddag$} \\
\\ {\it $^\star$ Department of Physics and Astronomy, Northwestern
University, Evanston,
Illinois 60208, USA} \\ {\it
$^\dagger$ Materials Science Division, Argonne National Laboratory,
Argonne, Illinois
60439, USA} \\ {\it
$^\ddag$  National High Magnetic Field Laboratory Tallahassee,
Florida 32310, USA} \\
\vspace{-5pt}

\noindent \dotfill\
\\ {\bf \noindent One of the puzzling aspects of high temperature
superconductors
is the prevalence of magnetism
in the normal state and the persistence of superconductivity in very high
magnetic fields.  Generally, superconductivity
and magnetism are not compatible.  But recent neutron scattering
results\cite{lake01} indicate that antiferromagnetism can appear
deep in the superconducting state in an applied magnetic field.
Magnetic fields penetrate a superconductor in the form of
quantized flux lines each one representing a vortex of supercurrents.
Superconductivity is suppressed in the core of the vortex
   and it has been suggested that antiferromagnetism might develop
there\cite{arovas97}.  To address this
question it is important to perform electronic structural studies
with spatial resolution.  Here we report on implementation of a high
field NMR imaging experiment\cite{takigawa99,wortis00,Morr00}
that allows spatial resolution of the electronic behavior both
inside and outside the vortex cores. Outside we find strong
antiferromagnetic fluctuations, and localized inside there are electronic states rather
different from those found in conventional superconductors.}

\begin{figure}[h]
\centerline{\epsfxsize0.90\hsize\epsffile{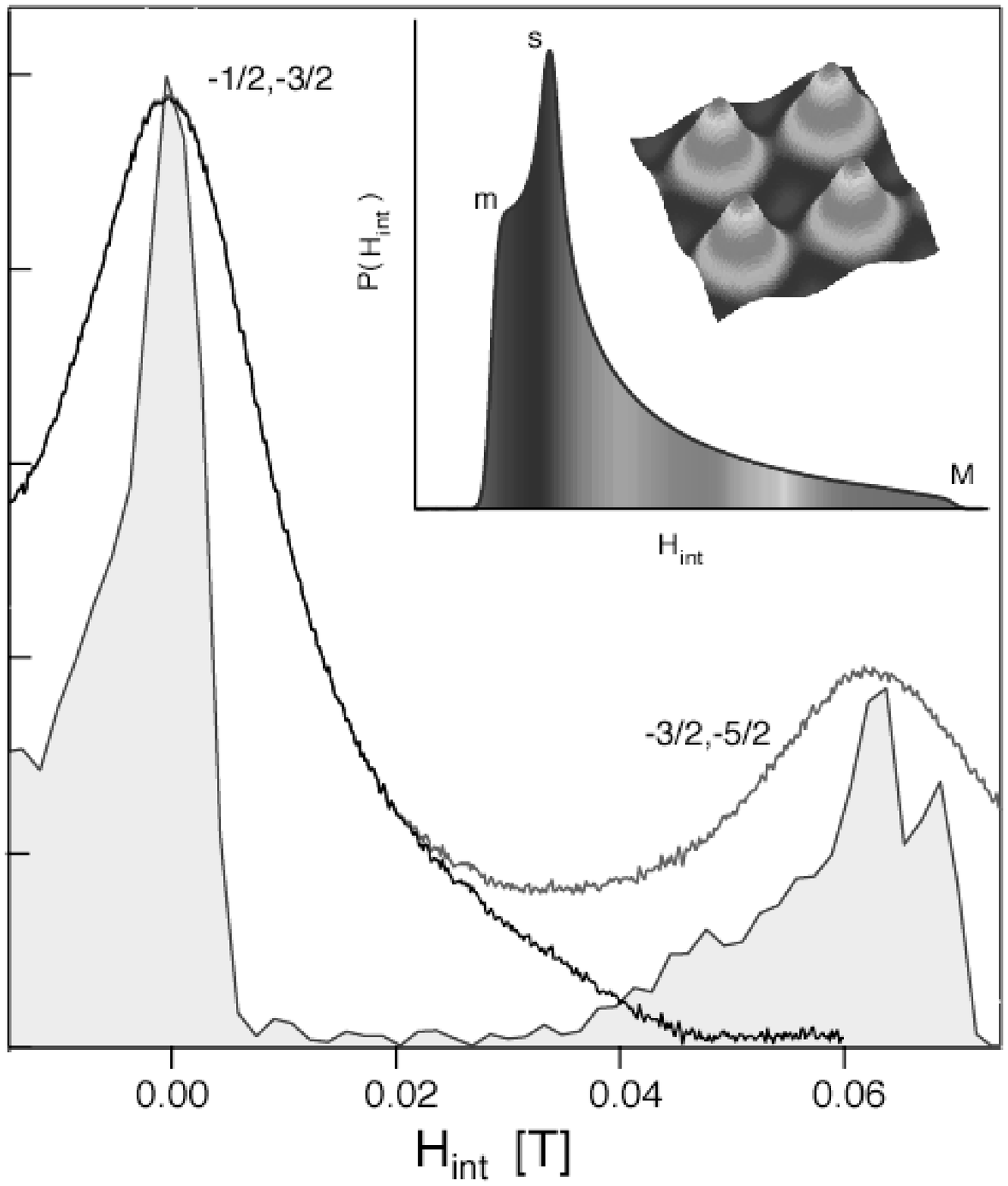}}
\begin{minipage}{0.93\hsize}
\caption[]{\label{Fig1}\small
    Planar $^{17}$O spectra versus the internal magnetic  field,
$H_{\scriptsize int}$, showing broadening from vortices.
$H_{\scriptsize int}={\omega  \mathord{\left/ {\vphantom
{\omega  {{}^{17}\gamma }}}
\right. \kern-\nulldelimiterspace} {{}^{17}\gamma }}-{\it H_0}$ where
$^{17}\gamma$ is the gyromagnetic ratio for oxygen and the
spectrometer frequency,
$\omega$, is set to resonance  at the peak frequency for the \mbox{-1/2
$\leftrightarrow$ -3/2} transition in the applied field ${\it
H_0}$=13 T at 11 K.     For clarity
only two quadupolar satellite resonances are shown.  In the normal
state  at 100 K (blue) there
is  a sharp edge which broadens at low temperature (red). The spectra
are shifted such that
internal fields are measured relative  to the peak.   The black trace
represents the \mbox{-1/2
$\leftrightarrow$ -3/2} part of the  spectrum obtained by subtraction
of the \mbox{-3/2
$\leftrightarrow$ -5/2} transition. The latter was obtained from the
convolution of the normal
state  spectrum  with a broadening function measured independently on
the \mbox{3/2
$\leftrightarrow$ 1/2} transition (not  shown here).  Inset: We have 
calculated \cite{Brandt97}   the internal magnetic field
profile and the corresponding field  distribution function for a
$80^\circ$ vortex lattice at ${\it H_0}$=13 T, with a penetration depth of $\lambda =
1500$ \AA. The  minimum field at the center of the vortex lattice unit cell (m), the
maximum field at the vortex cores (M), and the saddle-point field
midway between
vortices (s), are shown. }
\end{minipage}
\end{figure}

\begin{figure}[h]
\centerline{\epsfxsize0.95\hsize\epsffile{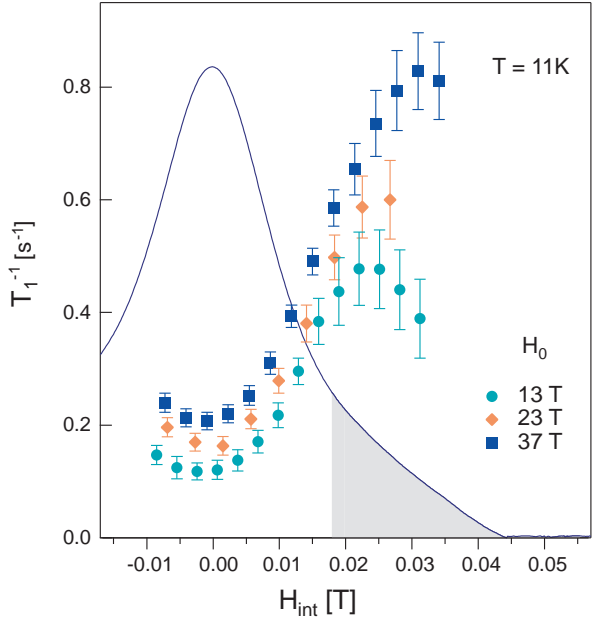}}
\begin{minipage}{0.93\hsize}
\caption[]{\label{Fig2}\small
    Planar $^{17}$O spin-lattice relaxation rate across  the vortex
lattice NMR spectrum  at 11
K.  The spectrum is the \mbox{-1/2 $\leftrightarrow$ -3/2} transition
at 37 T.  The shaded region is
the estimated area  occupied by the vortex cores at 37 T.  We used a
superconducting
magnet, 13 T; a Bitter-style electromagnet, 23 T;  and a
hybrid of the two technologies, 37 T. We obtained precise spectra
using  a field sweep technique and we measured the relaxation
rate  using progressive saturation\cite{mitrovic00}.}
\end{minipage}
\end{figure}

Supercurrents that form a vortex are expected to decay inversely with
distance from the
vortex core giving rise to a spatial distribution of internal
magnetic fields, $P(H_{int})$, shown  in \mbox{Fig. \ref{Fig1}}.
The color coding identifies spatial positions in the lattice of
vortices with their corresponding location
in the field  distribution for a vortex lattice. This field distribution
will be the same  as the NMR spectrum for
any nucleus with small intrinsic broadening as is the case
for $^{17}$O in YBCO.   Its narrow spectrum in the normal state (blue
trace) is shown  in the figure for two
of the four quadrupolar satellite resonances and compared with the
vortex broadened spectrum  (red trace)
in the  superconducting state \cite{reyes97,curro00}. We isolate  the $-1/2
\leftrightarrow -3/2$ transition giving the black trace,
corresponding to the expected field distribution
shown in the inset.   Consequently,  our NMR experiment allows a
spatially resolved study of the vortex structure. During the 
\noindent
NMR experiment  the oxygen nuclei are perturbed and then
relax back to thermal equilibrium at the spin-lattice relaxation rate,
which is determined by
the density of   electronic states  near the Fermi energy.
   We have made spin-lattice
relaxation rate  measurements, \mbox{Fig. \ref{Fig2}}, that are also spatially
resolved and we interpret our results in terms of the
electronic excitation spectrum both inside and outside the vortex cores.
Within the core we
expect  NMR to be sensitive to localized electronic states such as
have been observed in conventional
superconductors\cite{hess89} and recently in high temperature
superconductors\cite{MaggioAp95,pan2000}.
High magnetic fields allow us to use the Zeeman effect to study such
electronic excitations.
Varying the  
\noindent
field up to 37 T, we sweep  through a significant range
of corresponding
Zeeman energies, $\sim \pm 2.5$ meV, larger than the thermal energy.

Our aligned powder sample of YBa$_2$Cu$_3$O$_{7-\delta}$ is
near-optimally doped and
$\sim60\%$
$^{17}$O-enriched.  The crystal
$\hat  c$-axis was aligned with the direction of the applied
magnetic field. Low-field magnetization data show a sharp transition  at
$T_c(0)=92.5\,\mbox{K}$.
High magnetic fields are essential to increase sensitivity to vortex
cores by increasing
their fraction of the NMR spectrum.
In a field of 37 T vortices are
$\sim 80$ \AA $\,$ apart as compared with their core radius given
by the size of the electron-pair wavefunction,
\mbox{$\xi \approx 16$ \AA}, and so the cores occupy $\sim15
\%$ of both the total volume of the sample and the NMR spectrum.
We have shaded the corresponding region in \mbox{Fig.
\ref{Fig2}}.  The high field also suppresses spin diffusion and
vortex vibrations which complicate interpretation of the spin
lattice  relaxation at low field\cite{curro00}.

\begin{figure}[h]
\centerline{\epsfxsize0.65\hsize\epsffile{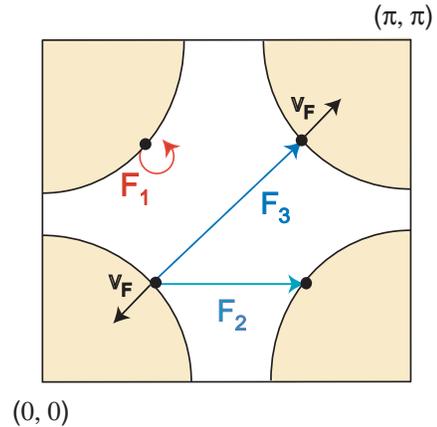}}
\vspace{8pt}
\begin{minipage}{0.93\hsize}
\caption[]{\label{FigFF}\small Sketch of the Fermi surface for YBCO and
the three processes that  contribute to the
NMR rate at low temperatures.  The three
processes, F$_1$, F$_2$, and  F$_3$,  correspond to the choice of nodes,
points where superconductivity is fully suppressed,
(black dots) for the initial and final
quasiparticle states. The spin lattice relaxation rate is given by,
$T_1^{-1} \sim \langle N_i(\epsilon ) N_f(\epsilon) \rangle$,
where $N_{\alpha}=\sum_{ {\bf k}} \delta (\epsilon -E_{{\bf k},\alpha })$
are the initial ($\alpha=i$) and final ($\alpha=f$) density of states
(DOS) respectively.
The sum over wavevector ${\bf k}$ is restricted to the regions around
the nodes,
${\bf k}_{\alpha }$, and the product of the DOS is integrated over
energy $\epsilon $ in the range $k_B T$ near the Fermi energy, $E_F$.
The excitation energies $E_{{\bf k},\alpha } $ for quasiparticles moving
in the superflow field with momentum ${\bf p}_s$
are Doppler shifted by an amount $D_{\alpha }={\bf v}_{F,\alpha} \cdot
{\bf p}_s$,
where the Fermi velocity at node ${\bf k}_{\alpha }$ is given in terms
of the normal state excitation energy $\epsilon_{{\bf k}}$ by,
${\bf v}_{F,\alpha}=
\partial \epsilon_{{\bf k}}/\partial
{\bf k}|_{{\bf k}={\bf k}_{\alpha}}$.
Note that for process F$_3$ the relation $D_i=-D_f$ holds, in contrast to,
for example, process F$_1$ where $D_i=D_f$.
In sufficiently high magnetic field the Zeeman energy, becomes important
   and
enters initial and final states with different signs ($Z_{i,f}=\pm Z$),
leading to,
$E_{{\bf k},\alpha }  =\sqrt{\epsilon_{{\bf k}}^2+\Delta_{{\bf k}}^2}
- Z_{\alpha } + D_{\alpha }$.
Since the DOS depends linearly on energy near the nodes,
the NMR rate is proportional to the product given by  \mbox{Eq.
\ref{eqT1Energy}}.}
\end{minipage}
\end{figure}

For high-temperature superconductors the gap in the excitation spectrum,
$\Delta _{\bf k}$,   is believed to have $d$-wave symmetry, implying
   that there are 4 nodes  on the Fermi surface where the gap vanishes.
At low temperature the electronic excitations,
called quasiparticles, come  exclusively from the nodal regions shown in
\mbox{Fig. \ref{FigFF}}.
They can be probed with  NMR spin lattice relaxation, with the rate given by,

\noindent
\begin{equation}
\label{eqT1Energy}
T_1^{-1} \sim \langle |\epsilon -Z +D_i|\; |\epsilon +Z +D_f| \rangle ,
\end{equation}
where $\epsilon $ is  of the order of $k_B T$; $Z=-\frac{1}{2}\gamma_e
\hbar H_0$, is the Zeeman energy;
$D_{\alpha }={\bf v}_{F,\alpha} \cdot {\bf p}_s$ is the Doppler
shift, and $\alpha=i,f$
corresponds to initial and final electronic states near the nodes.
Depending on the relative magnitude of $k_BT$, $Z$, and $D$ the
dependence of the relaxation rate on temperature and field will show
different behavior for each of three processes shown in \mbox{Fig.
\ref{FigFF}}.
We can vary the first two terms, $k_BT$, $Z$, by changing temperature
and magnetic field.  The Doppler term is
proportional to the vortex currents, and varies with internal field,
$H_{int}$, across the NMR spectrum. This term is a microscopic
analogue of the classical frequency shift of a wave in a moving reference
frame, the latter being associated with the superconducting condensate.

At the peak of the NMR spectrum,  $H_{int} = 0$ in \mbox{Fig. \ref{Fig2}},
we observe that the rate  increases as a function of the external field,
${\it H_0}$.     This dependence comes from the Zeeman interaction.
The Doppler term is small in this region since
   the supercurrents cancel at the saddle-point.
As we increase the applied field the Zeeman
interaction dominates the energy spectrum near the nodes, i.e. $Z$,
  equivalent to 25 K at 37 T, is larger than
$k_B T$ at  11 K, and from \mbox{Eq. \ref{eqT1Energy}} the NMR
rate rises quadratically with field as $\langle |{Z}^2 -{\epsilon}^2|\rangle$.

To the right of the peak in the spectrum toward the vortex core,  \mbox{Fig.
\ref{Fig2}}, we find that the  rate increases by a factor of $\sim 4 -
5$. Here the superflow momentum  increases  sufficiently that
the  Doppler term in \mbox{Eq. \ref{eqT1Energy}}
   becomes large compared to both $Z$ and
$k_BT$ and it dominates the quasiparticle energy.
Our  observations show that
$T_1^{-1}$, and consequently the density of electronic states (DOS),
depends on distance, $r$, from the core.  In  particular,
we expect the rate to increase as
$T_1^{-1}\sim \left( {p_s}\right)^2 \sim r^{-2}$.  The enhancement of
$T_1^{-1}$ depends only on $r$ and should  not be dependent on the
applied magnetic field other than the Zeeman
interaction.  This is not to be confused with Volovik's
prediction\cite{Volovik93} that the  average
DOS, $\left\langle {N(0)} \right\rangle \sim \sqrt {\it H_0}$, which
corresponds to $\left\langle
T_1^{-1}\right\rangle\sim {\it H_0} $ln${\it H_0}$ where the average
is over the entire spectrum. For conventional
superconductors without nodes, the Doppler term has a negligible
effect.  Our observation of the Doppler shift extending to
the saddle-point region shows that, even  in  high magnetic fields, the
superconducting gap has well defined nodes.

Outside the vortex core we find that the effect of
applied  magnetic field from 13 to 37 T is simply to displace the
relaxation rate data on a vertical scale in \mbox{Fig. \ref{Fig2}}.
This is possible if the
Doppler term  in \mbox{Eq. \ref{eqT1Energy}}   changes sign for
initial and final states,
just like the Zeeman term, and can only
occur if those quasiparticle  states are not from the same node.
Then, at low temperatures, the rate is $\sim  {D}^2 + {Z}^2$
and increases  quadratically with field.  Otherwise, initial and
final states come from the same  node and the rate is $\sim  {D}^2 - {Z}^2$,
having the wrong field dependence in the regime $D > Z$. We have calculated
the relative strengths of the three possible processes and their dependence on
magnetic field.  The calculation is performed  using the known geometric form
factors and the imaginary part of the electronic susceptibility
which we extract by fitting the data. To account for our
observed field dependence, the spin susceptibility must be strongly
peaked near a wavevector corresponding to the F$_2$ or F$_3$ process,
at least a factor of 7 greater than
that at zero wavevector.  This indicates that  strong
antiferromagnetic correlations between quasiparticles
coexist  with superconductivity at low temperatures.
Antiferromagnetically correlated quasiparticles  were identified by
neutron scattering\cite{lake01} in LSCO, a related compound, and are now
confirmed in YBCO from NMR.  Lake {\it et al.}
speculated that if there was antiferromagnetism in the vortex core
it might polarize the  intervening medium and account for
their experiment.

   Near the vortex core, the NMR rate decreases after reaching a
maximum  in the applied field of
13 T  as shown 
\begin{figure}[h]
\centerline{\epsfxsize0.95\hsize\epsffile{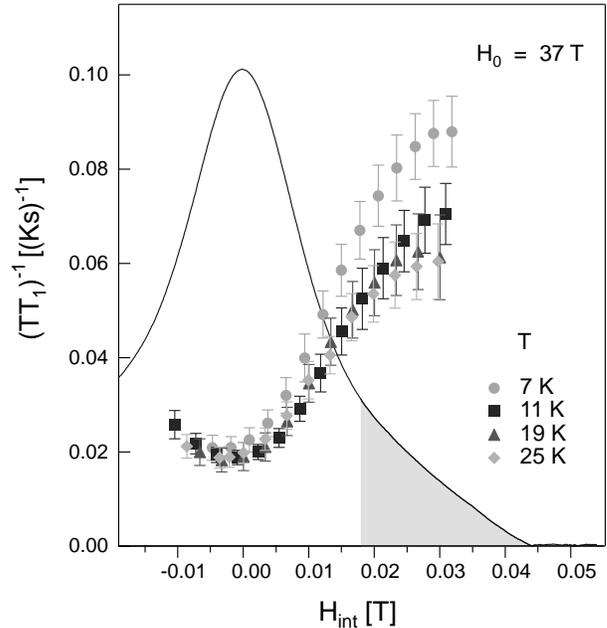}}
\begin{minipage}{0.93\hsize}
\caption[]{\label{Fig3}\small Spin-lattice relaxation rate of planar
$^{17}$O divided by temperature  as a function of internal magnetic
field.  The shaded region
is the area  occupied by  the vortex cores at 37 T.  For a $d$-wave
superconductor  in the
temperature range where
$k_BT$ is much less than $\gamma_e \hbar {\it H_0}$,  $(T_1T)^{-1}$   is
expected to be constant.  We  clearly observe this behavior near the
peak in the NMR spectrum. }
\end{minipage}
\end{figure}
\noindent
in \mbox{Fig. \ref{Fig2}}, reflecting a drop-off in
$p_s$. At higher applied fields the rate increases
significantly, well beyond  its maximum  value at 13 T.  This
increase with field exceeds that
measured outside the cores by a factor of four and is  most
pronounced at the lowest
temperatures,  \mbox{Fig. \ref{Fig3}}.   We attribute  this
substantial  excess rate to
vortex core states.   When superconductivity is suppressed near surfaces
or in the vortex core, localized electronic
states, called Andreev bound states, can appear\cite{hess89,les92}.
Their signature is a peak in the density of states at the Fermi
energy, called a zero-bias anomaly in tunneling  experiments. Similar
behavior might be  expected in
the vortex core of a high temperature superconductor.  If we
interpret our relaxation rate in  the same way
as for  quasiparticles outside the cores, following the discussion in
the caption to  \mbox{Fig. \ref{FigFF}} and \mbox{Eq.
\ref{eqT1Energy}}, then both field and temperature dependence are opposite
to what is required for the zero-bias anomaly.
Consequently, instead of a peak in the density of states there is a
mini-gap, $\sim
\pm 5$ meV, much sharper  than the variation of the DOS outside the
vortex core.
Previous reports of vortex core states have been  based
on  charge tunneling in YBCO\cite{MaggioAp95} and BSSCO\cite{pan2000}.
Near $E_F$, $\sim
\pm 5.5$ meV for  YBCO,  an added structure in the DOS from the
vortex core region
separated by a minigap was observed.
This surprising
result is inconsistent with theoretical predictions \cite{franz98}.
Furthermore, tunneling is
extremely sensitive to the surface and depends on an understanding  of the
tunneling matrix elements \cite{wu00}. In contrast
NMR probes bulk material.  And yet, both experiments can have a
similar interpretation.

NMR is directly sensitive to the magnetic character of electronic excitations.
   So an alternate explanation of our results
for the vortex core region might involve antiferromagnetism as predicted
\cite{arovas97} from a SO(5) nonlinear $\sigma$ model. In this case
field induced
antiferromagnetic excitations could strongly
enhance the NMR rate consistent with our observations.  In future
work, spatially
resolved NMR relaxation  experiments can be extended to understand
the effects of doping
on vortex core states and possibly identify a connection with the
spin-pseudogap found in the normal state.

\vspace{-0.2cm}
\bibliographystyle{unsrt}

\vspace{0.3cm}

{\bf \noindent Acknowledgements \\ \\}  This work was supported by
the Science and Technology Center for
Superconductivity, the Materials Research Center at Northwestern
University, and the National  High
Magnetic Field Laboratory supported by the National Science
Foundation and the State of Florida.   ME
acknowledges support from U.S. Dept. of Energy, Office of Science. We
are grateful to J. A. Sauls,
   J. Moreno, R. Wortis  and K. Machida for helpful discussions.
\\
\\ Correspondence and requests for materials should be addressed  to  W.P.H.
(e-mail:w-halperin@northwestern.edu).

\end{multicols}

\end{document}